# LTE SFBC MIMO Transmitter Modelling and Performance Evaluation

Gabriela Morillo
School of Computer Science and IT
University College Cork
Cork, Ireland
g.morillo@cs.ucc.ie

John Cosmas
College of Engineering, Design and Physical Sciences
Brunel University London
Uxbridge, United Kingdom
john.cosmas@brunel.ac.uk

*Abstract*— High data rates are one of the most prevalent requirements in current mobile communications. To cover this and other high standards regarding performance, increasing coverage, capacity, and reliability, numerous works have proposed the development of systems employing the combination of several techniques such as Multiple Input Multiple Output (MIMO) wireless technologies with Orthogonal Frequency Division Multiplexing (OFDM) in the evolving 4G wireless communications. Our proposed system is based on the 2x2 MIMO antenna technique, which is defined to enhance the performance of radio communication systems in terms of capacity and spectral efficiency, and the OFDM technique, which can be implemented using two types of sub-carrier mapping modes: Space-Time Block Coding and Space Frequency Block Code. SFBC has been considered in our developed model. The main advantage of SFBC over STBC is that SFBC encodes two modulated symbols over two subcarriers of the same OFDM symbol, whereas STBC encodes two modulated symbols over two subcarriers of the same OFDM symbol; thus, the coding is performed in the frequency domain. Our solution aims to demonstrate the performance analysis of the Space Frequency Block Codes scheme, increasing the Signal Noise Ratio (SNR) at the receiver and decreasing the Bit Error Rate (BER) through the use of 4 QAM, 16 QAM and 64QAM modulation over a 2x2 MIMO channel for an LTE downlink transmission, in different channel radio environments.

In this work, an analytical tool to evaluate the performance of SFBC - Orthogonal Frequency Division Multiplexing, using two transmit antennas and two receive antennas has been implemented, and the analysis using the average SNR has been considered as a sufficient statistic to describe the performance of SFBC in the 3GPP Long Term Evolution system over Multiple Input Multiple Output channels.

## Introduction

THE wireless industry has evolved in subscribers and technologies in the last few years. The evolution of wireless access technologies allows better performance and high efficiency in mobile communications through the improvement of the functionalities such as mobile voice, capacity, coverage, and access to advanced mobile services with a wide range of data rates according to the multiuser requirements (Kumar et al., 2010).

Mobile cellular networks have been categorised into generations. The last most used: the fourth generation (4G), tries to achieve new levels of user experience and multi-services capacity, integrating the existing mobile technologies and protocols.

The 3rd Generation Partnership Project (3GPP) has standardised these new data networks to increase the speed and capacity of the wireless communications systems. As was mentioned previously, the last widely used standard is the 4G Long Term Evolution (LTE), which is an evolution of the current second and third-generation wireless networks, and it was developed to achieve the high user requirements regarding data rate, speed, costs, and expansion of providing services.

To cover these demands, LTE combines the most recent schemes and technologies, such as Orthogonal Frequency Division Multiplexing (OFDM) and Multiple-Input Multiple-Output (MIMO) techniques. Furthermore, LTE in the downlink transmissions uses OFDMA (Orthogonal Frequency Division Multiple Access), which allows the encoding of digital data on multiple carrier frequencies.

According to Torabi and Conan (2014), a group of flat fading subchannels are converted from the wideband frequency selective channel by OFDMA, features that made OFDMA an efficient method to improve the spectral efficiency of wireless communications. Nevertheless, OFDM signals allow adding frequency domain scheduling to time domain scheduling, making this technique more resistant to frequency selective fading than a single carrier system.

Long Term Evolution Technology was designed to evolve the radio access technology and considers that all services should be packet switched instead of circuit switching, which is the model of the previous technologies (Jemmali, 2013). According to Serra (2013), LTE also was designed to increase data rates, improve spectrum efficiency, and allow spectrum flexibility. Furthermore, for real-time services such as VoIP, LTE reduces the packet latency, the cost regarding the radio access network and the migration from earlier 3GPP releases and one of the main advantages of the implementation of this technology is the fact that it can simplify the network converting it into an all-IP-packet-based network architecture in which the functionalities are terminated in the base stations (evolved Node B).

Finally, to achieve the required data throughput and reliability, the LTE standard uses MIMO, which is a powerful technique to improve the performance of a wireless communication system.

This project aims to determine an efficient model in *Matlab* to implement an LTE MIMO transmitter using an SFBC coding scheme and to evaluate its performance.

## SFBC MIMO IN LTE

Several recent studies investigating Long Term Evolution technology have been carried out and recognised that the combination of Multiple-Input Multiple-Output (MIMO) wireless technology with the Orthogonal Frequency Division Multiplexing (OFDM) technique allows high data rate, large capacity, robustness to multipath fading and high performance in the 4G wireless communications (May-Win & Naing, 2014).

The high requirements regarding reliable and fast communications in the channel demand a technology which can cover these challenges. As it is known, 4G can achieve speeds up to ten times higher than 3G, while OFDM can mitigate the multipath effects and prevent the frequency selective fading and the Inter Symbol Interference (ISI). Furthermore, in this project, MIMO with the Space Frequency Block Coding has been used to achieve high data transmissions over wireless links through multiple antennas because they guarantee greatly improved performance in wireless communication systems.

Early publications into the Long-Term Evolution- LTE specifications defined by the 3GPP emphasise that the required parameters for high-quality communication include a peak data rate for a downlink of 300 Mbps or more within 20 MHz of bandwidth using MIMO-OFDM. According to Heng and Louay (2015), there are two alternative ways in which using MIMO can be achieved improved performance in LTE; the first alternative includes the spatial multiplexing transmission mode, which allows capacity gains. The second alternative includes transmit diversity, which uses the SFBC to obtain diversity gains.

Much of the current literature regarding the transmit diversity schemes put particular attention to the Space Time Block Coding scheme. Although STBC is the most popular scheme because it is used in 3G networks, it assumes a perfect and quasi-static channel (Horvat et al., 2010) and the coding technique is performed across the number of OFDM symbols, which are equivalent to a number of transmit antennas (Molinero, 2013). This represents the major limitation of this scheme because of its channel variation sensitivity in the time domain, which represents performance degradation owing to the channel difference over the block duration (Kalbat and Al-Dweik, 2015). In contrast, the 3GPP LTE standard implements the SFBC scheme, in which the coding is performed across the subcarriers within the interval of the OFDM symbol, where it presents a good BER performance in time-varying channels.

### LTE Downlink Reference Signal Structure

The reference signal has been defined in the LTE downlink transmissions to determine the channel estimation. The resource elements allocated in the time-frequency domain are responsible for carrying the reference signal sequence for a specific cell.

The resource elements result from the combination of one OFDM symbol in the time domain and one subcarrier in the frequency domain. This structure is presented in Fig. 1, where RE represents the Resource Element, RB the Resource Block and RS the Reference Signal.

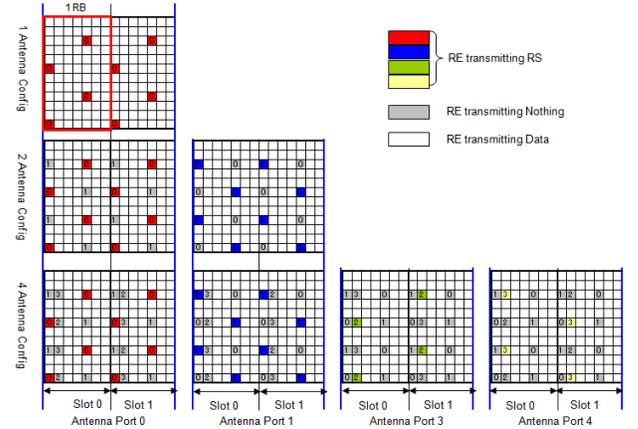

Fig. 1. Downlink cell-specific reference signals in LTE

In the frequency domain, the resources are grouped in sets of 12 subcarriers, which means that they use 180 MHz with a subcarrier spacing of 15 KHz; 12 subcarriers form a resource block.

The Resource Element (RE) is the smallest unit of resource, and it is formed by one subcarrier with a duration of one OFDM symbol. Furthermore, a Resource Block consists of 48 Resource Elements when the normal cyclic prefix length is used; and 72 Resource Elements when the extended cyclic prefix length is used.

### OFDMA in the Downlink

In LTE downlink transmissions, the system can classify or tag the data across a standard number of subcarriers according to the resource blocks into which they are split. Independent of the LTE signal bandwidth, the resource block in LTE involves 12 subcarriers and covers one slot in the time frame; as a result, different numbers of resource blocks for each LTE signal bandwidth have been obtained, as presented in Table I.

TABLE I
BANDWIDTH AND RESOURCES BLOCKS IN LTE (POOLE, 2015)

| Channel Bandwidth (MHz) | 1.4 | 3 | 5 | 10 | 15 | 20 |
|---|---|---|---|---|---|---|
| Number of resource blocks | 6 | 15 | 25 | 50 | 75 | 100 |

### MIMO in LTE

This section presents MIMO technology only to extend its applicability in the LTE transmission modes.

Multi-antennas are implemented to improve the data rates,

capacity, and spectral efficiency of a radio communication system. Since wireless communications have been in a continuously growing race with fixed-line communications regarding certain parameters such as speed, throughput and capacity, mobile networks present a continuous evolution (Uyoata and Noras, 2014). One of these innovations is the inclusion of MIMO technology in data transmission to achieve an increase in data rates.

A limitation in mobile communications is known as fading, which consists of the rapid changing of the amplitude of the propagated signal, and to overcome its effects, LTE uses Multiple Input Multiple Output, which uses more than one antenna in both transmitter and receiver. In general terms, a MIMO system consists of m transmit antennas and n receive antennas, as is shown in Fig. 2. The signal received is called y, and it is the result of the expression (1):

$$y = H * x \qquad (1)$$

Where; H is the transmission matrix which contains the channel impulse responses $h_{nm}$ of the channel between the transmit and receive antennas. The matrix order defines the number of independent data streams which can be transmitted simultaneously. According to how the data is transmitted, two modes in which MIMO can improve LTE system performance have been defined:

Spatial Diversity: The same data is transmitted redundantly over more than one transmit antenna. STBC and SFBC are used to increase the robustness of data transmission.

Spatial Multiplexing: The data is transmitted in separate streams, and they are transmitted simultaneously over the same air interface resources. Several parameters must be considered in the transmission, such as the definition of the pilot or reference signals and the channel estimation for each signal on the transmitter side.

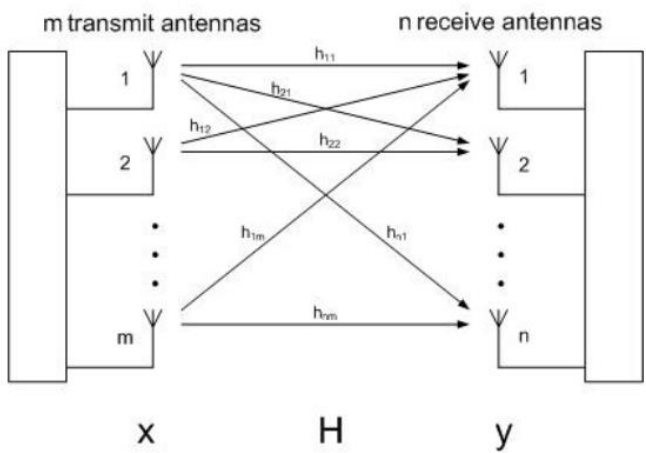

Fig. 2. MIMO System with m transmit antennas and n receive antennas [Schulz, 2015]

*Transmit Diversity Schemes in LTE Downlink*

The transmit diversity scheme is a technique that allows increasing diversity and coverage, especially if it is applied in terminals with bad propagation conditions (Ciochina et al., 2008). Furthermore, in mobile systems, the implementation of transmit diversity is common in the downlink because it is cheaper and easier to install multiple antennas at the eNodeB than install multiple antennas at every portable device (Iordache and Marghescu, 2013).

According to Schulz (2015), the transmit diversity sends the same information through several antennas, in which each antenna stream uses different coding and frequency resources to improve the signal-to-noise ratio (SNR) parameter and the transmission.

In this section, two transmit diversity schemes are presented. The first is the Space Frequency Block Coding using two transmit antennas, and the second is the Frequency Switched Transmit Diversity using four transmit antennas.

*Space Frequency Block Coding*

The diversity scheme in LTE is called SFBC if the physical channel uses two eNodeB antennas to transmit diversity operation. Fig. 3 illustrates the main principle of the SFBC transmission in 4G networks; it can be observed that in the SFBC model, a pair of consecutive modulated symbols called $(X_i, X_{i+1}; X_{i+2}, X_{i+3}; \ldots; X_{i+n-1}, X_{i+n})$ are mapped directly in the first antenna port as consecutive samples in time; while in the second port, a pair of consecutive swapped and transformed symbols called $(-X^*_{i+1}, X^*_i; -X^*_{i+3}, X^*_{i+2}; \ldots; -X^*_{i+n}, X^*_{i+n-1})$ are mapped in time, allowing the consecutive vectors on the different antennas to be orthogonal (Zarrinkoub, 2014).

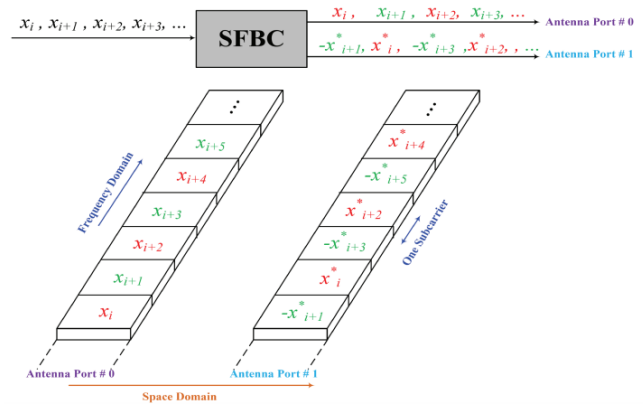

Fig. 3: Space Frequency Block Codes in LTE [Torabi et al., 2014]

The main advantage of SFBC compared with STBC is that Space Frequency Block Coding applies coding across the subcarriers in the OFDM symbol interval, while Space Time Block Coding is done across the number of OFDM symbols which are equivalent to the number of transmit antennas.

In practice, Space Frequency Block Coding diversity scheme is the implementation in the frequency domain of the STBC technique defined by Alamouti; Fig. 4 presents a comparison between a 2x2 STBC scheme and an SFBC scheme. In (a), the STBC with Alamouti code transmits a pair of adjacent, no consecutive, symbols in the time domain symbols formed by a modulated and a conjugated symbol called $(S_1, -S_2^*)$; and in the following sample time, a pair of swapped and transformed symbols called $(S_2, S_1^*)$ are transmitted.

By contrast, in (b), a pair of consecutive modulated symbols called $(S_1, S_2)$ is mapped directly in the first antenna port; while in the second port, a pair of consecutive swapped and transformed symbols called $(-S_2^*, S_1^*)$ are mapped in time.

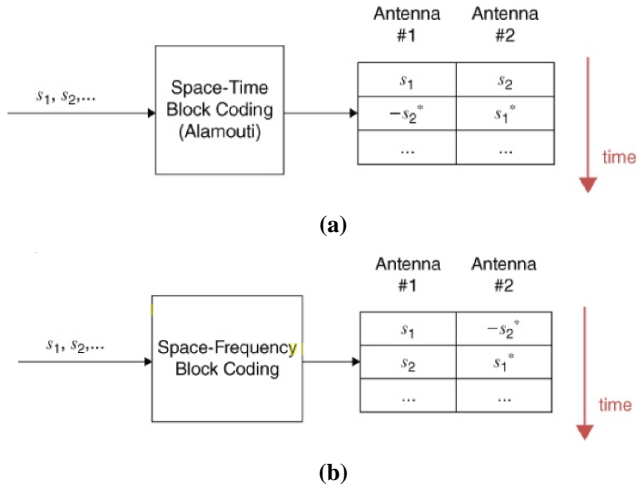

Fig. 4. Comparison between (a) 2X2 STBC and (b) 2X2 SFBC schemes

SFBC operates on a pair of complex-valued modulation symbols, and it is known that in LTE, the number of available OFDM symbols in a sub-frame is often odd; SFBC allows each pair of modulation symbols to be mapped in the OFDM subcarrier of the first antenna directly, besides, for the second antenna is considered a reverse order: complex conjugated and signed reversed. On the receiver side, the mobile unit must be notified regarding the SFBC transmission and at the receiver, a linear operation should be applied to the signal.

The principle of 4X4 SFBC transmission is shown in Fig.5. on the left, the bit stream to be transmitted can be observed, and next in the middle section, the conversion (coding process) is a performance in a Block, one by one bit. The OFDM symbols in the subcarrier are separated by Space between each antenna, and the subcarriers are separated by Frequency between subcarriers.

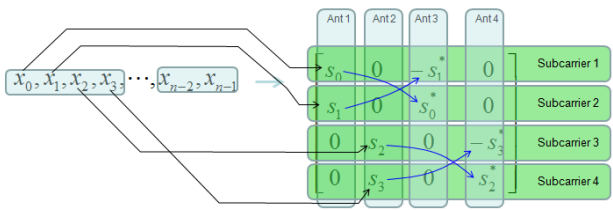

Fig. 5. Space Frequency Block Code Scheme in LTE

*Frequency Switched Transmit Diversity*

When four transmit antennas have been used, the diversity scheme is called Switched Transmit Diversity. In Fig. 6., Fig the main principle of operation of the FSTD scheme is presented, in which in two of the four antennas, a pair of modulated symbols are transmitted using the SFBC scheme, while the other two antennas are not transmitting.

In FSTD, a pair of transmit antennas switch the transmission at each frequency slot; as can be observed, in the first frequency slot, the first two pairs of symbols $(X_i, X_{i+1})$ and $(X_i^*, -X_{i+1}^*)$ are transmitted through *Antenna Port #0* and *Antenna Port #2*, respectively, while any symbol is transmitted through *Antenna Port #1* and *Antenna Port #3*. For the second frequency slot, the next two pairs of symbols are transmitted $(X_{i+2}, X_{i+3})$ and $(-X_{i+3}^*, X_{i+2}^*)$ through *Antenna Port #1* and *Antenna Port #3*, while any symbol is transmitted through *Antenna Port #0* and *Antenna Port #2*.

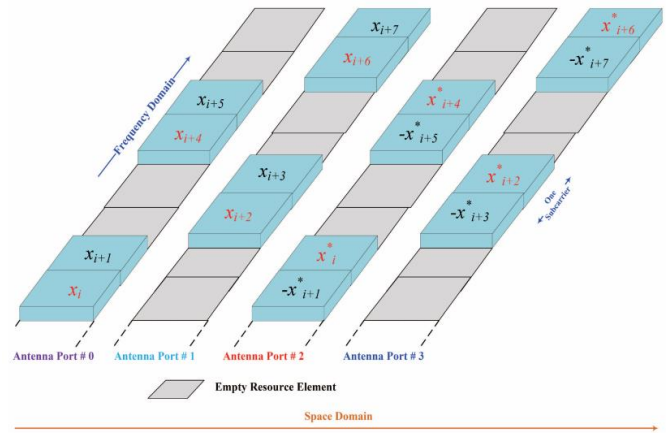

Fig. 6. Frequency Switched Transmit Diversity scheme [Torabi et al., 2014]

SFBC TRANSMITTER MODELLING AND PERFORMANCE EVALUATION

The details of the implementation of the LTE SFBC 2x2 MIMO antenna transmitter, the description of the main functions of the code, the BER simulation results and the corresponding analysis have been presented in this section.

*Methodology*

The development of a real-time communication system requires an extensive range of resources, skills and time. Design requirements such as Modularity which allows modifications, extensions and even reuse of several parts of the code; and Downscale Specifications that allows focus in the core algorithms and the critical aspects of the overall system architecture has been considered for the model implementation.

Furthermore, the implemented model is based on the LTE Release 8 Physical Downlink and Uplink Layer Transmission System created by Madina Olzhabayera at Brunel University London.

*Model Design*

The fundamental criterion applied throughout the design, implementation and validation phases is a multi-stage testing strategy, as shown in Fig.7. Briefly, the model starts from the Data Generation stage, which performs a sequential insertion of symbols in the initial scenario. It continues with the encoding stages, which include the phases between the Modulation and the Inverse Fast Fourier Transform, then the signals pass through the MIMO and the AWGN channel, and finally, the decoding process is executed in the following stages, between the Fast Fourier Transform and the Measuring BER stage.

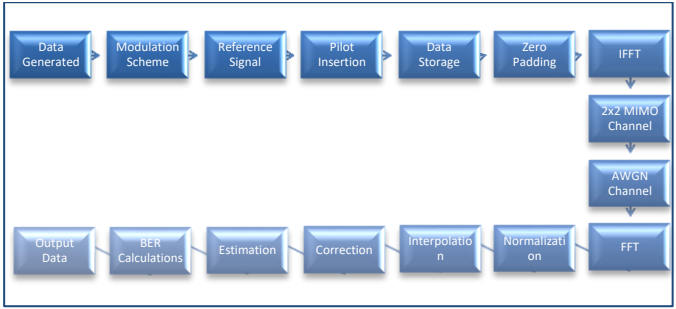

Fig, 7. Model Design LTE SFBC 2x2 MIMO antennas [Olzhabayera, 2015]

*Design Criteria*

*1) Basic Transmitter Modeling*

The first essential requirement for the implementation of the model is the definition of the transmitted signal, which includes the definition of the OFDM parameters, the frame format, the frame length, the duplexing mode, the location and value of the pilots, the guard-band and bandwidth sizes. The principal stages on the transmitter side are presented below.

*Data Generate*

Considering that $N$ represents the number of symbols as input on the transmitter side, the random data can be generated in two different ways. In the first alternative, the data is generated consecutively as a burst of symbols, which means that they are mapped directly in the first antenna port as consecutive samples in time, as shown in Fig. 8.

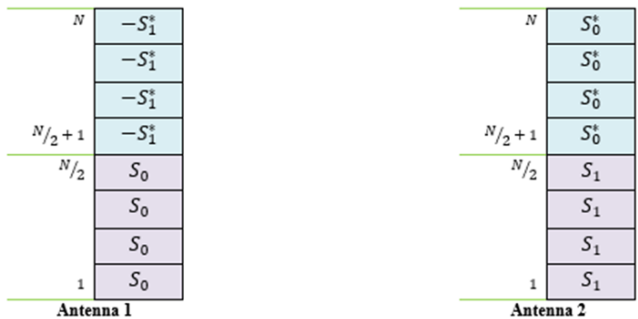

Fig. 8. Consecutively Symbol Insertion

In the second alternative, the data is generated through the interleaving of symbols, as illustrated in Fig. 9.

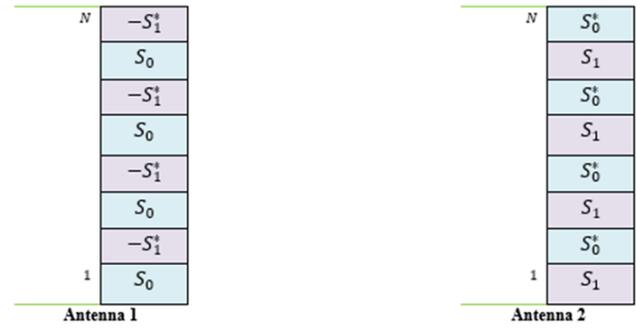

Fig. 9. Consecutively Symbol Insertion

In this project, the first alternative has been selected for the implemented model.

The location of the Reference Signal in the First Transmission and the Conjugate Transmission for both Antenna 1 and Antenna 2, using Space Frequency Block Code in LTE downlink, is presented in Table II.

TABLE II
LOCATION OF THE REFERENCE SIGNAL USING SFBC

[Table II: A grid showing the location of Reference Signals across Antenna 1 and Antenna 2 for First Transmission and Conjugate Transmission, with subcarrier indices $-S_{1,1}^*$ through $-S_{1,7}^*$ and $S_{0,1}$ through $S_{0,7}$ for Antenna 1, and $S_{1,1}$ through $S_{1,7}$ for Antenna 2. Reference signals $R_0$ and $R_1$ are placed at specific positions, with $X$ markers indicating null positions.]

*Modulation Scheme*

The blocks of the received digital data have been mapped into data blocks using three types of modulation techniques: 4-QAM, 16-QAM and 64-QAM depending on the required speed.

At this stage the following considerations were made: the data is allocated in the Resources Blocks, each RB is a segment of the OFDM spectrum which consists of 12 subcarriers of 15 $kHz$ for a total of 180 $kHz$. As can be observed in Fig. 10., in the time domain, there are seven segments per subcarrier which have a duration of 5 $ms$.

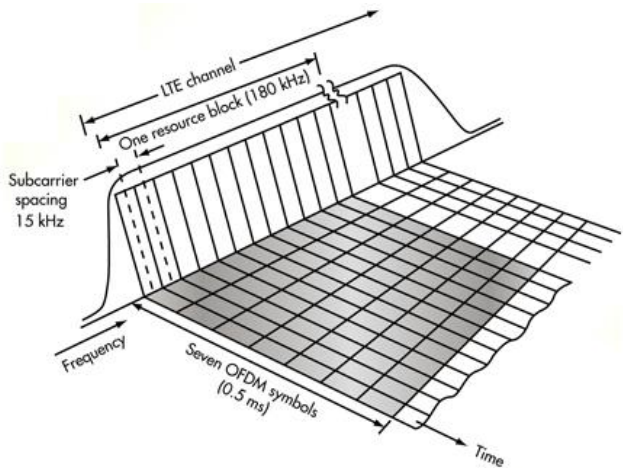

Fig. 10. Design Criteria for the Data Transmission [Frenzel, 2013)

*Reference Signal*

In the frequency domain, LTE allows to define the channel-dependent scheduling, which means that to perform downlink scheduling, the mobile terminal can provide the Channel State Information (CSI) to the base station, measured in the reference signal transmitted (Zarrinkoub, 2014).

The reference signals are generated with synchronised sequence generators in both the transmitter and receiver.

*Pilot Insertion*

To facilitate the estimation of the channel characteristics, the pilot symbols have been inserted in unique positions to ensure no interference between them and to guarantee a reliable estimate of complex gains in each resource element.

*Data Storage*

In the 2x2 MIMO antennas model, two data inputs have been defined: $a_1$ and $a_2$, wherein the lower side of the spectral model:
- $a_1$ is the $S_0$ symbol.
- $a_2$ is the negative conjugate of $S_1$ symbol.

These arrays of information were used on the upper side of the spectral model as follows in (1) and (2):
- $a_{1-\text{upper}}(N/2 + k) = a_2^*(k)$  (1)
- $a_{2-\text{upper}}(N/2 + k) = -a_1^*(k)$  (2)

*Zero Padding*

LTE uses the Fast Fourier Transform to divide the channel into many slighter channels with lower data rates, also transforms the signals between time and frequency domains and achieves orthogonality to each other through the generation of coefficients. These coefficients depend on the available bandwidth; commonly, this number is bigger than the size of the number of symbols in the mapped subcarrier; for this reason, the zero padding is added to the end of a time-domain signal to increase its length and to achieve that the signals have a power-of-two number of samples.

*IFFT*

According to Zarrinkoub (2014), the input bits are modulated in the transmitter and the Inverse Fast Fourier Transform-IFFT is applied to the modulated symbols before the channel modelling processes. On the receiver side, the inverse operation is done to the demodulation.

*Channel Estimation*

The channel modelling is performed by the combination of two channels, a MIMO channel with an Additive White Gaussian Noise -AWGN channel.

*MIMO Channel*

In the model, the main configuration made in the MIMO Channel was regarding parameters such as the antenna configurations, multipath delay, maximum Doppler shifts and spatial correlation levels.

*AWGN Channel*

In the model, the noise added before the OFDM will arise with the Fast Fourier Transform; to normalise the SNR at the receiver, the AWGN channel has been defined so that the noise is scaled; for that a specific and equal amount of white noise is added in every frequency spectrum.

*2) Basic Receiver Modeling*

*FFT*

The OFMD symbols are demodulated in this block. The Fast Fourier Transform is applied to the two antennas and to obtain symbols with the same size of the number of the subcarrier, the zero padding has been removed from the demodulated data.

*Normalisation*

The pilot's normalisation process is applied in each individual frame, in order to change the amplitude and phase of all the pilots to amplitude of 1 and phase 0; this because initially the pilots are received with amplitudes of 1 but phases of 45°, 135°, -45° and -135°.

*Interpolation*

The Interpolation is performed between pilot's estimates to create a channel estimated for all Resource Elements. This block has been defined in order to estimate the subcarriers frequency response (Olzhabayera, 2015); the interpolation of the pilots in frequency is made in the vertical direction, while the interpolation of the pilots in time is made in the horizontal direction.

For SFBC coding scheme, the allocation of the pilot subcarriers is made only in some frequencies but continuously in the time as is depicted in Fig. 11, and for the rest of frequencies, the channel frequency response is obtained through the calculation of the frequency response of the data subcarriers.

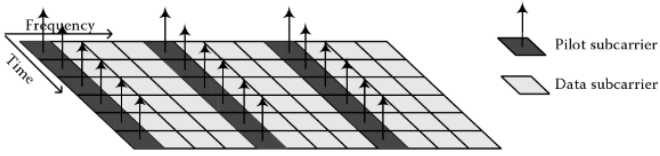

Fig. 11. Time continuous – frequency spaced pilot allocation [Furht and Ahson, 2009]

The channel frequency response is defined in (3):
$$y = Hx + n \qquad (3)$$
Where; $y$ is the received signal, $x$ are the transmitted symbols, $H$ is the channel coefficient matrix, and $n$ is the Gaussian Noise.

*Correction and Estimation*

In a 2x2 MIMO antenna system working with SFBC as a scheme for received data correction, each pair of modulated OFDM symbols is mapped, as illustrated in Fig 12. It can be observed that the bit stream to be transmitted passes through the coding process which is performed in Block, then the OFDM symbols in the subcarrier are separated by Space between each antenna, and the subcarriers are separated by Frequency between subcarriers.

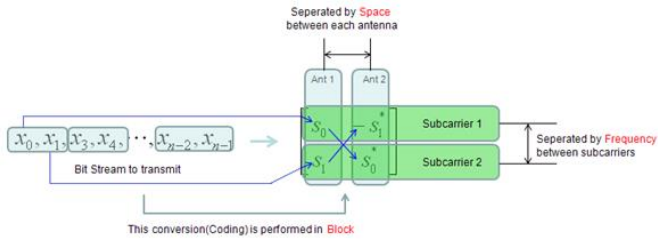

Fig. 12. SFBC 2x2 MIMO antennas

Therefore, the mapping of each pair of OFMD symbols in each antenna is performed as shown in Table III:

TABLE III
OFDM SYMBOLS ARRANGEMENT

| Antenna | Frequency/Time | $t_0$ | $t_1$ |
|---|---|---|---|
| $ANT_0$ | $f_0$ | $x_o$ | |
|  | $f_1$ | $-x_1^*$ | |
| $ANT_1$ | $f_0$ | $x_1$ | |
|  | $f_1$ | $x_0^*$ | |

Then, the $x$ matrix is defined as:
$$\begin{matrix} & f_0 & f_1 \\ Antenna_0 - t_o & \\ Antenna_1 - t_1 & \end{matrix} \begin{bmatrix} x_o & -x_1^* \\ x_1 & x_0^* \end{bmatrix}$$

To establish the estimated data for each antenna, firstly, the channel configuration presented in Fig. 13 has been used to define the equations (4), (5), (6) and (7) for the receiver signal.

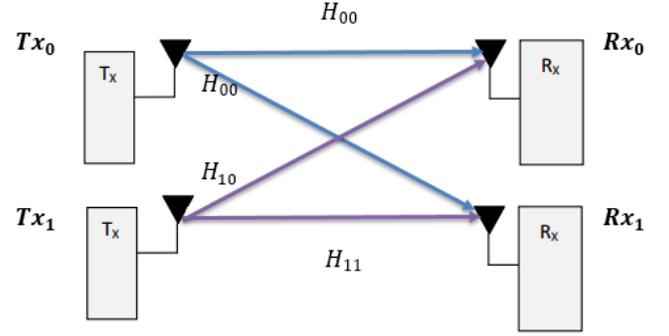

Fig. 13. 2x2 MIMO antennas – Channel Configuration

For Antenna 0:
$$y_{00} = x_0 H_{oo} + x_1 H_{10} \qquad (4)$$
$$y_{01} = x_0 H_{o1} + x_1 H_{11} \qquad (5)$$

For Antenna 1:
$$y_{10} = x_0^* H_{10} - x_1^* H_{00} \qquad (6)$$
$$y_{11} = x_0^* H_{11} - x_1^* H_{01} \qquad (7)$$

As it can be demonstrated, the equations remain the same as that in the Space Time Block Coding Scheme analysed in the model created at Brunel University London by Madina Olzhabayera; hereby, the channel response for each antenna remains equal, as well as the estimated data for each antenna which is defined as in (8) and (9).

$$\hat{x}_0 = \frac{H_{00}^* y_{00} + H_{10} y_{10}^* + H_{01}^* y_{01} + H_{11} y_{11}^*}{|H_{00}|^2 + |H_{10}|^2 + |H_{01}|^2 + |H_{11}|^2} \qquad (8)$$

$$\hat{x}_1 = \frac{H_{10}^* y_{00} - H_{00} y_{10}^* + H_{11}^* y_{01} - H_{01} y_{11}^*}{|H_{00}|^2 + |H_{10}|^2 + |H_{01}|^2 + |H_{11}|^2} \qquad (9)$$

*BER Calculations*

Firstly, the received symbols must be converted into bits and then the Matlab function `biterr` which computes the number of bit errors.

*Developed Code*

The following considerations have been made:

− In the 2x2 MIMO antennas model, the data outputs defined as $a_1$ and $a_2$ are the result of the combination of two flags: Negative Flag and Conjugate Flag, and its value depend of the QAM modulation scheme.

− For the Reference Signal and Pilot Insertion, the functions `PilotSFBCInsertDown` and `PilotSFBCUpperInsertDown` have been created to cover the design criteria defined for the insertion of the reference signals and pilots, respectively.

Fig. 14 depicts the symbol distribution in the spectral model; in order to optimise the code, the lower side of the spectral model has been implemented through the `PilotSFBCInsertDown` function, and the upper side of the spectral model has been defined in the `PilotSFBCUpperInsertDown` function.

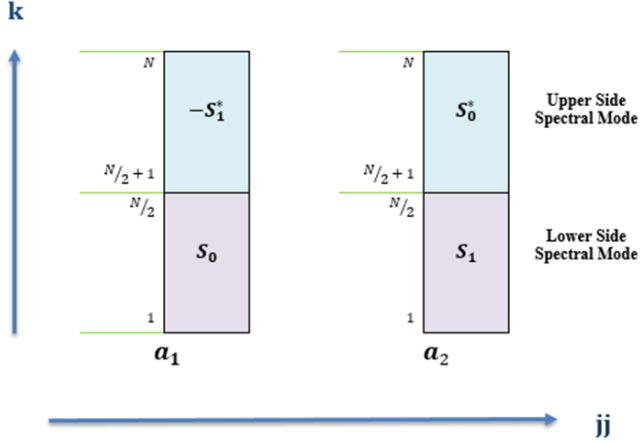

Fig. 14. Symbol Distribution in the Spectral Model

The insertion of the $S_0$ symbol is defined according to the different types of modulation, where the limits in the frequency domain ranging from 1 to N/2, also the Negative Flag and the Conjugate Flag have been set with zero. The resulting vector is called $a_1$.

The insertion of the $S_1$ symbol is defied with the limits in the frequency domain ranging from 1 to $N/2$, and the Negative Flag and the Conjugate Flag have been set as 0. The resulting vector is called $a_2$.

The outputs $a_1$ and $a_2$ obtained above have been used to insert the data on the upper side of the spectral model.

Then, the symbol's magnitudes are switched between each limited range. First, the insertion of the $-S_1^*$ symbol is defined according to the different types of modulation, whereas the limits in the frequency domain ranging from 1 to $N/2$; also, the definition of the vector has been made according to the equation (1), and the negative conjugate operation has been applied to the vector $a_2$ in the $N/2 + k$ frequency range.

The insertion of the $S_0^*$ symbol is defined for the different types of modulation, whereas the limits in the frequency domain ranging from 1 to $N/2$ to. Alike, the definition of the vector has been made according to equation (2), and the conjugate operation has been applied to the vector $a_1$ in the $N/2 + k$ frequency range.

- On the Decoder side, the mathematical model defined in (8) and (9) has been applied for the Correction and Estimation phase.
The limits in the time domain have been defined between 1 to 14 according to the slots of time, while the frequency slots have been established from 1 to $N$.
Also, it can be verified that the frequency is variable in a value of $f + N$; where $N$ is defined as (10).

$$\frac{length(All\_received\_antenna1\_data)}{2} \quad (10)$$

RESULTS AND DISCUSSION

The performance evaluation of the LTE SFBC MIMO system has been analysed through the definition of two scenarios. The first includes a variation of the QAM order, and the second consists of a variation of the radio channel.

*By Varying the QAM Order*

The value of the Signal Noise Ratio will be adapted in phases to achieve a perfect signal transmission, which means $BER = 0\ dB$. The settings parameters presented in Table IV remain constant for the analysis of each modulation.

In Fig 15, it can be verified that BER of zero has been obtained of around 9 dB, 22 dB and 33 dB for 4 QAM, 16 QAM and 64 QAM, respectively. Therefore, it can be concluded that with an increase in QAM order, the BER performance degrades as the number of errors in the received signal increase.

Moreover, it can also be observed that when the QAM order increases, a higher SNR is obtained; this allows better performance because the number of sub-channels increases, which means that high data rates, less distortion and fewer retransmissions can be obtained.

TABLE IV
SETTING PARAMETERS FOR DIFFERENT QAM ORDERS

| Parameters | | Value |
|---|---|---|
| Transmission Mode | Mode | SFBC 2x2 Downlink |
| Antennas | Tx Antenna Cross-Correlation | 0.5 |
| | Rx Antenna Cross-Correlation | 0.5 |
| Frame Structure Selection | Mode System | FDD |
| | TDD Configuration Options | 0 |
| | Resources Blocks | 6 |
| Bandwidth Selection | Bandwidth (MHz) | 1.4 |
| | DFT Size (M) | 36 |
| | System Frequency (GHz) | 2.7 |
| Symbol Selection | Structure | Frame |
| | Number Frames | 4 |
| Channel Multipath | Radio Channel Type | Rayleigh |
| | Radio Environment | User Defined |
| | Radio Channel K Factor | 1000 |
| | Speed of Mobile (Km/h) | 3 |

Furthermore, for an LTE SFBC 2x2 antennas MIMO, assuming a User Defined environment, the average BER performance at a high Signal to Noise Ratio decreases until to reach zero bits with errors. Therefore, assuming this kind of scenario, it is possible to compute a perfect received signal.

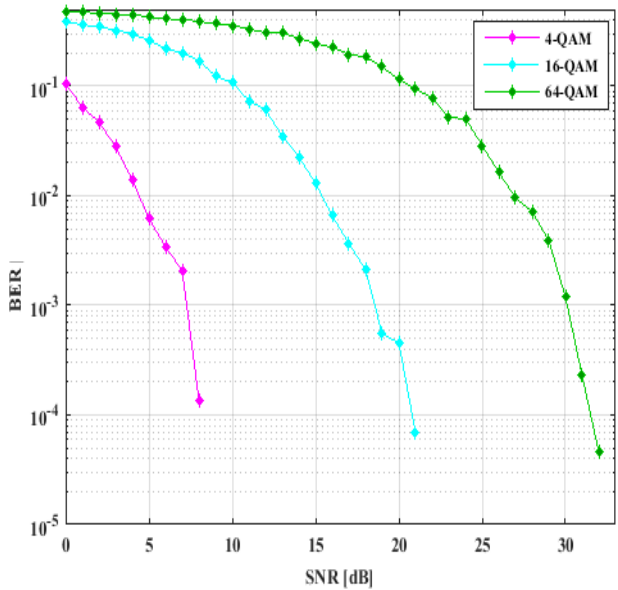

Fig. 15. SNR vs. BER for 4 QAM, 16 QAM and 64 QAM Modulations for User-Defined Channel

*Varying the Radio Environment*

The implemented model defines six different radio environments. The Signal Noise Ratio value will be adapted to achieve a perfect signal transmission, which means BER=0. The simulation has been performed for the three modulation types and each radio environment.

The settings parameters presented in Table V remain constant for the analysis in each scenario.

TABLE V
SETTING PARAMETERS FOR THE DIFFERENT RADIO ENVIRONMENTS

| Parameters | | Value |
|---|---|---|
| Transmission Mode | Mode | SFBC 2x2 Downlink |
| Antennas | Tx Antenna Cross Correlation | 0.5 |
|  | Rx Antenna Cross Correlation | 0.5 |
| Frame Structure Selection | Mode System | FDD |
|  | TDD Configuration Options | 0 |
|  | Resources Blocks | 6 |
| Bandwidth Selection | Bandwidth (MHz) | 1.4 |
|  | DFT Size (M) | 36 |
|  | System Frequency (GHz) | 2.7 |
| Symbol Selection | Structure | Frame |
|  | Number Frames | 4 |
| Channel Multipath | Radio Channel Type | Rayleigh |
|  | Radio Environment | Variable |
|  | Radio Channel K Factor | 1000 |
|  | Speed of Mobile (Km/h) | 3 |

The results shown in the sections below consider the two relevant cases. The first case is when the received signal is performed with $SNR = 0\ dB$, and the second is when the received signal reaches $BER = 0$ or a value close to zero.

*1) AWGN*

From Fig. 16, it can be concluded that with an increase in QAM order, the BER performance degrades, and a higher SNR is obtained, which represents a better performance because high data rates, less distortion and fewer retransmissions are required.

Furthermore, it can be stated that for an LTE SFBC - 2x2 antenna MIMO, assuming an AWGN environment, the average BER performance at a high Signal Noise Ratio decreases until to reach zero bits with errors. Therefore, assuming this kind of scenario, it is possible to compute a perfect received signal.

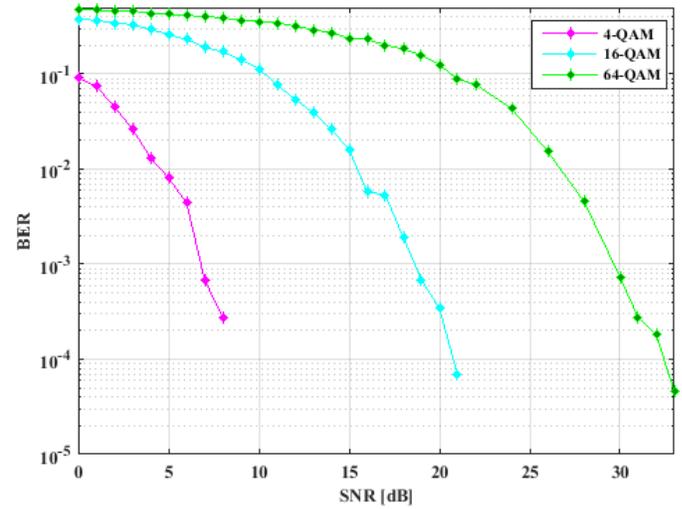

Fig. 16. Comparison M-QAM Modulation for AWGN Environment

*2) Rural Area*

The same trend as the previous environments has been identified; when the QAM order increases, the BER performance degrades, and the SNR increases, allowing better performance. Furthermore, as it is shown in Fig. 17., the average BER performance at high SNR decreases until to reach zero bits with errors only for the 4 QAM modulation; therefore, assuming this kind of scenario, it is possible to compute a perfect received signal only for 4 QAM; the other schemes are more sensitive to non-ideal factors, and a BER value close to zero has been achieved.

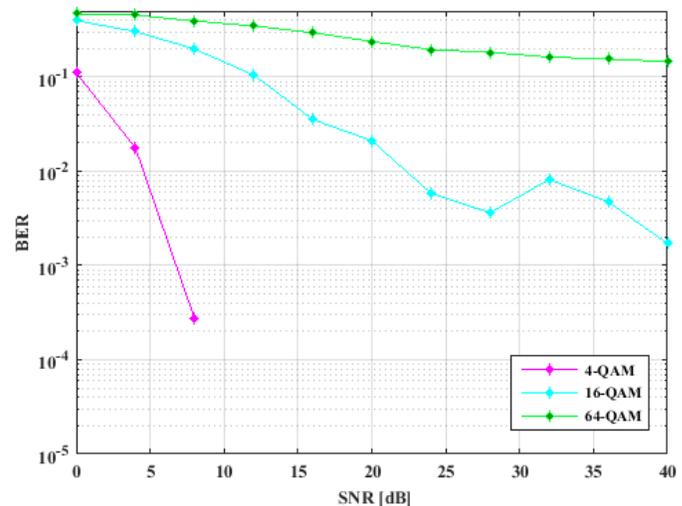

Fig. 17. Comparison M-QAM Modulation for Rural Area Environment

*3) Typical Urban*

Because of the longer multipath delays, this environment is more complicated than the previous one. From Fig. 18, it can be verified that a BER of zero is difficult to achieve. The expected trend BER was obtained until SNR= 20 $dB$, and the BER values decreased gradually between SNR= 0 $dB$ and SNR= 20 $dB$; then, they slightly fluctuated for the following SNR values. Thus, it can be stated that the BER performance remains constant after a specific value of SNR. Moreover, it can also be observed that when the QAM order increases, the constant value of BER is higher in each modulation scheme. Therefore, it can be concluded that for an LTE SFBC - 2x2 antenna MIMO, assuming a Typical Urban environment, it is difficult to accurately determine the number of errors with the conditions of that environment.

*4) Bad Urban*

Like the Typical Urban environment, zero BER is difficult to achieve. The expected trend BER was obtained until SNR= 20 $dB$. Then, it slightly fluctuates for the next SNR values, as can be observed in Fig. 19. Thus, the BER performance remains stable after a specific value of SNR.

*5) Hilly Terrain.*

One of the most unfavourable conditions has been recognised for this scenario. The BER of zero is difficult to achieve, and as the Typical and Bad Urban environments, the BER value decreases gradually between SNR= 0 $dB$ and SNR= 20 $dB$; then, it slightly fluctuates for the next SNR values, as it is shown in Fig. 20.

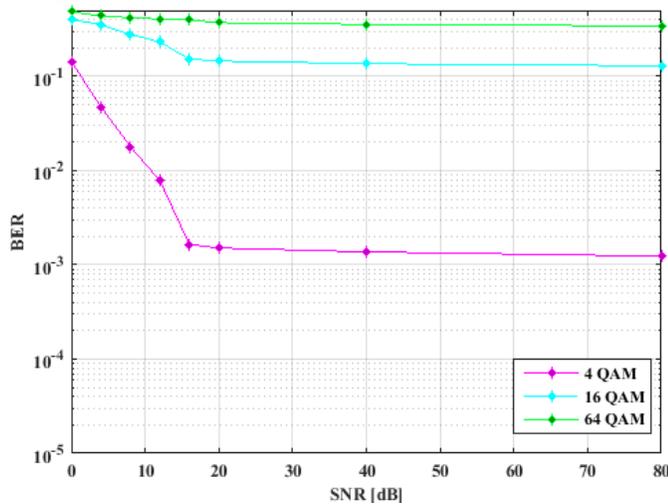

Fig. 18. Comparison of M-QAM Modulation for Typical Urban Environment

## CONCLUSIONS

The purpose of the current research was to determine the impact of the Space Frequency Block Codes in LTE downlink transmissions. Firstly, this project provides a comprehensive review of the literature on the available 3GPP standard to understand the physical layer of an LTE system and its performance. Furthermore, the second major outcome was the fully functional implemented Space Frequency Block Code (SFBC) of 2x2 antennas MIMO for LTE model, which allows to determine and analyse of the Bit Error Rate and the Signal to Noise Ratio at various scenarios in a MIMO channel.

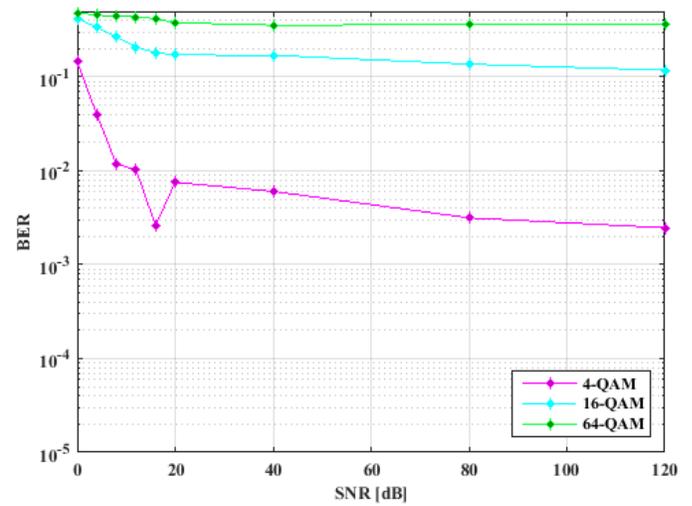

Fig. 19. Comparison of M-QAM Modulation for Bad Urban Environment

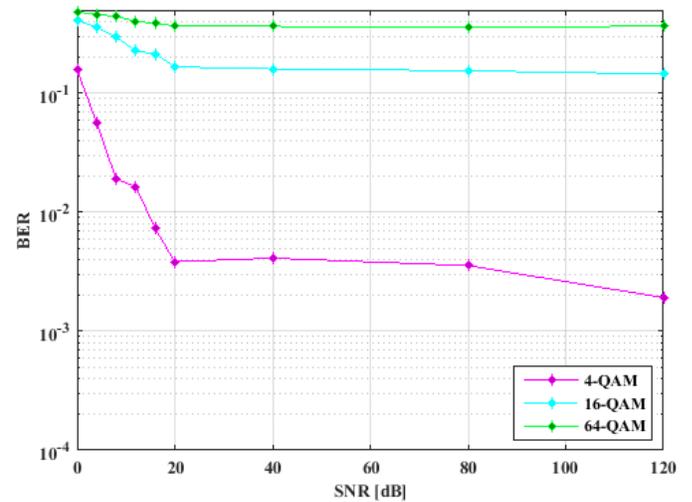

Fig. 20. Comparison M-QAM Modulation for Hilly Terrain Environment

The average BER assessed in terms of Signal to Noise Radio was used as a performance metric to evaluate the effect of transmit diversity scheme chosen in the LTE downlink transmission. The performance evaluation results of the LTE SFBC MIMO model obtained from the mathematical expressions and design criteria derived in this project were analysed by defining two scenarios.

The first scenario includes a variation of the QAM order under a unique environment: User Defined where the average BER performance at high SNR decreases until to reach zero bits with errors; therefore, it can be concluded that assuming this kind of scenario, it is possible to compute a perfect received signal.

The second scenario includes a variation of the radio channel. The obtained results allow us to conclude that the lowest BER value is obtained in the AWGN, User Defined and Rural Area environments, while the performance of the received signal decline in the Typical Urban, Bad Urban and Hilly Terrain environments because these scenarios present a higher level of complication due to factors as the longer multipath or power delays. Therefore, the number of bits with errors is higher than in the previous three scenarios.

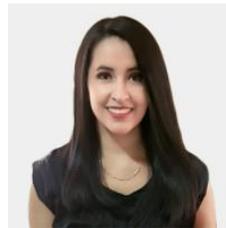

**Morillo Gabriela**. received her MSc. in Computer Communication Networks at Brunel University London in the United Kingdom and her Engineering in Electronics and Network Information at The National Polytechnic School in Ecuador.

Her research interests include the development of new, disruptive, and sustainable solutions at the lowest environmental and economic cost. At SFI ADVANCE CRT, Centre for Research Training in Advanced Networks for Sustainable Societies, she focuses on the security issues for IoT that will contribute to safeguarding the end-users sensitive information, avoiding attacks of capacity and data theft, and allowing location privacy and data security.

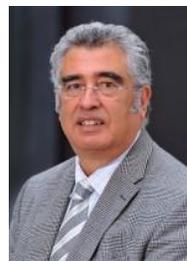

**Cosmas John.** (M'86) received his BEng in Electronic Engineering at Liverpool University, UK, in 1978 and a PhD in Image Processing at Imperial College, University of London, UK, in 1986. He worked for five years in the industry, first with Tube Investments and then with Fairchild Camera and Instruments. After completing his PhD, he worked for 13 years as a lecturer in digital systems and telecommunications at Queen Mary College, University of London. Since 1999, he has worked for Brunel University, first as a reader and then in 2002 as a Professor of multimedia systems. His current research interests are multimedia broadcast communications systems, which evolved from his longstanding interests in video codecs (MPEG 4/7), and mobile communication systems (DECT/GSM/GPRS/UMTS). Prof. Cosmas currently serves as associate editor for IEEE Trans. Broadcasting. He also contributes to the Digital TV Group's "Mobile Applications" sub-group and to the Digital Video Broadcast's Technical Module: Converged Broadcast and Mobile Services (CBMS).